# Energy-efficient techniques for combating the influence of reactive jamming using Non-Orthogonal Multiple Access and Distributed Antenna Systems

Joumana Farah[(1)], Jacques Akiki[(1)], Eric Pierre Simon[(2)]

[(1)] Department of Electricity and Electronics, Faculty of Engineering,
Lebanese University, Roumieh, Lebanon
[(2)] Institute of Electronics, Microelectronics and Nanotechnology (IEMN) Lab,
University of Lille, Villeneuve d'Ascq 59650, France

*Abstract*—The aim of this work is to propose new approaches for maximizing the energy efficiency of downlink 5G mobile communication systems, in the presence of a reactive jammer. The concepts of non-orthogonal multiple access (NOMA) and distributed antenna systems (DAS) are exploited to devise joint subband, power and antenna assignment techniques, so as to guarantee a certain quality of service (QoS) to users. Also, the scheduler relies on jamming statistics, observed at the end of each timeslot, to perform resource allocation based on the prediction of the jammer behavior over the next timeslot. A particular care is given, in the proposed techniques, to maintain a moderate complexity at the receiver level, and to limit the number of active RRHs (remote radio heads) in the cell. Simulation results show that a proper combination of NOMA with DAS can allow a significant enhancement of the system robustness to jamming, with respect to centralized antenna systems and orthogonal multiple access.

## I. INTRODUCTION

The security of wireless communications has been lately one of the major concerns to the scientific community, especially with the advent of the fifth generation of communication networks. Unlike wired communications, wireless transmissions are very susceptible to jamming because of the shared wireless medium. When an ongoing legitimate communication is subject to jamming, it experiences a drop in its Quality of Service (QoS), which can even lead to Denial of Service (DoS) in intense cases, depending on the jammer capabilities. However, security is not the only issue that concerns mobile operators. Reducing their capital expenditure is also a major challenge that can only be achieved by efficiently exploiting the available resources: spectrum and energy. Moreover, the ever growing need for higher data rates to enable bandwidth-hungry mobile applications (web browsing, video conferencing, etc.) raises the required transmit power, which in turn increases electromagnetic radiations and $CO_2$ emissions.

The purpose of this study is to devise new resource allocation (RA) techniques that can enhance the robustness of 5G mobile systems against jammer attacks, while optimizing their energy efficiency. The originality of this work resides in the exploitation of two recent concepts to counteract the influence of jamming: distributed antenna systems (DAS) and non-orthogonal multiple access (NOMA).

The concept of DAS [1-4] consists of deploying the base station antennas in a distributed manner in the cell, instead of having multiple antennas installed on a single tower at the cell center. The remote units, called remote radio heads (RRH), are connected to a baseband unit (BBU) by fiber optics links. By reducing the average distance between users and antennas, the overall transmit power, necessary to ensure a certain quality of reception, is reduced in comparison to centralized antenna systems (CAS). Therefore, DAS can greatly reduce local electromagnetic radiation and $CO_2$ emissions and improve the system robustness to fading, interferences, shadowing, and path loss. The optimization of energy efficiency (EE) in the DAS context was the target of the works in [5] and [6]. In [5], proportional fairness scheduling is considered for subband allocation. In [6], subband assignment and power allocation (PA) are done in two separate stages. PA is performed by maximizing EE under the constraints of total transmit power per RRH, target Bit Error Rate and proportional fairness among users. The problem of power minimization in DAS was considered in [4] with a joint allocation of subcarrier, RRH and power to users, so as to guarantee user target rates. However, none of these studies have accounted for the presence of a jammer in the system.

In [7], a joint PA and scheduling technique was proposed for maximizing user throughputs in uplink CAS jammed transmissions. The case of a reactive jammer was considered and jamming history was observed to constitute statistics for predicting the jammer behavior. Also, three modes were used for uplink transmit power: conservatory, aggressive and exploratory. A PA strategy was also proposed in [8] for optimizing the power assignment of the legitimate user and the jammer over multiple channels so as to maximize the number of received packets, while achieving the Nash equilibrium of a zero-sum game. However, both studies focus on uplink transmission, and their extension to the downlink case and to different objectives (e.g. power minimization or EE maximization) is not straightforward.

NOMA has recently been introduced in 5G systems and consists on multiplexing several users on the same subband in the power domain, by taking advantage of the channel gain difference between users [9-12]. At the receiver side, user separation is done using successive interference cancellation (SIC). Compared to Orthogonal Multiple Access (OMA), NOMA allows a significant increase in the system throughput, or, alternatively, a great reduction of the transmit power [12].

To the best of our knowledge, no previous work has tackled the problem of alleviating the influence of jamming through the combination of NOMA and DAS techniques, while maximizing the system EE. This is the main contribution of this paper, which also includes the derivation of the SIC conditions in the presence of jamming. This paper is organized as follows: in Section II, we start by a detailed description of the system model and then develop the NOMA pairing conditions in the presence of jamming. Then, in Section III, we present several algorithmic solutions for the EE maximization problem, for OMA and NOMA transmission. Section IV presents a performance analysis of the different allocation strategies, while Section V concludes the paper.

## II. DESCRIPTION OF THE NOMA-DAS SYSTEM UNDER JAMMING ATTACK

In this section, we start by describing the communication system using the concepts of DAS and NOMA. Then, we derive the proper conditions for user pairing in NOMA under jamming attack, and formulate the system requirements.

The downlink cellular system consists of $R$ transmitting RRHs and $K$ mobile users randomly deployed over a cell. RRHs and users are equipped with a single antenna. Users transmit channel state information (CSI) to RRHs, and the BBU collects all CSI from RRHs. Then, it allocates subbands, powers and RRHs to users in such a way to guarantee a transmission rate of $R_{k,target}$ [bps] for each user $k$. The system bandwidth $B$ is equally divided into $S$ subbands. On the $n^{th}$ subband ($1 \leq n \leq S$), a maximum of $m(n)$ users $\{k_1, k_2, …, k_{m(n)}\}$ are chosen from the set of $K$ users, to be collocated on $n$ (or paired on $n$ when $m(n) = 2$). OMA transmission corresponds to the special case of $m(n) = 1$. The framework is schematized in Fig 1, where RRH $r$ powers the signal of user $k$ on subband $n$. A reactive jammer $J$ continuously monitors transmissions on all subbands. It only jams subband signals when the received signal power at the level of $J$ is higher than a threshold $P_{th}$. The jamming signal is supposed to have a duration equal to that of the legitimate signals, i.e., the timeslot duration. Also, the power detection performed by the jammer is supposed to be instantaneous at the beginning of each timeslot.

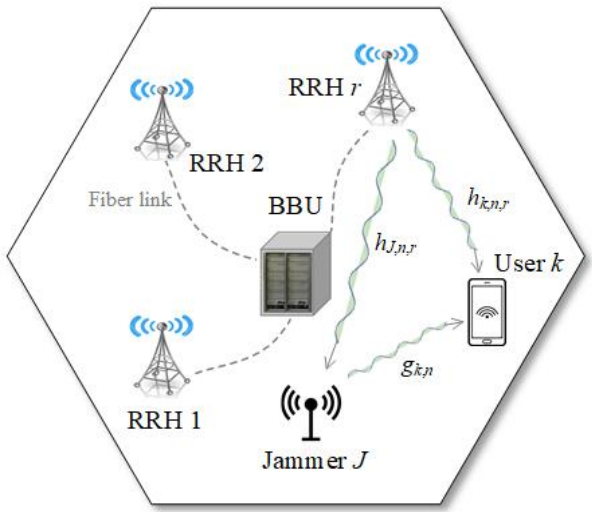

Fig. 1. NOMA-DAS transmission system in the presence of a jammer

Table 1 summarizes the main variables and parameters used in this study.

TABLE. 1. Definition of main variables and parameters

| Symbol | Significance |
|---|---|
| $B$ | System bandwidth |
| $S$ | Total number of subbands |
| $K$ | Number of users |
| $R$ | Number of RRHs |
| $g_{k,n}$ | Channel coefficient between Jammer $J$ and user $k$ on subband $n$ |
| $h_{k,n,r}$ | Channel coefficient between user $k$ and RRH $r$ on subband $n$ |
| $h_{J,n,r}$ | Channel coefficient between jammer $J$ and RRH $r$ on subband $n$ |
| $N_0$ | Noise power spectral density |
| $\sigma^2$ | Noise power ($\sigma^2 = N_0 B/S$) |
| $P_J$ | Jammer transmit power |
| $R_{k,n,r}$ | Rate of user $k$ on subband $n$ powered by RRH $r$ |
| $P_{k,n,r}$ | Power allocated for user $k$ on subband $n$ powered by RRH $r$ |
| $\alpha_{n,r}$ | Jammer triggering function on subband $n$ powered by RRH $r$ |

As in [7] that considers reactive jamming, it is assumed that the state of all users (positions in the cell and channel gain coefficients) as well as the jammer's remain constant during a time horizon $H_o$, taken as an integer number of timeslots. The BBU has no knowledge of the jammer triggering threshold $P_{th}$, nor of the jammer channel coefficients with respect to the RRHs ($h_{J,n,r}$). However, through an exchange of control information between users and RRHs, the BBU gathers information of the jammer powers on subbands where the jammer was triggered. In other words, the users periodically send information of their achieved rates or RSSI (received signal strength indicator), which allows the BBU to deduce if the jammer was triggered or not at the end of each timeslot, and estimate the values of the jamming power levels perceived by the users: $P_J g_{k,n}^2$. Through proper channel allocation techniques, the BBU can also estimate the channel coefficients between users and RRHs ($h_{k,n,r}$) assumed to be perfectly known in this work.

Also, while performing resource allocation (subband and power assignment), the BBU does not know beforehand whether the jammer will be triggered or not at the end of the current timeslot. For this reason, at the beginning of each timeslot, the BBU relies on the triggering outcome of previous timeslots in $H_o$. Such statistics are gathered as was done in [7] for CAS uplink transmission: the BBU keeps track, over the horizon $H_o$, and for each subband-RRH pair ($n$, $r$), of the RRH transmit power values that triggered the jammer and those that did not. Let:

- $P_{n,r}^1$ the largest transmit power on ($n$, $r$) that does not trigger the jammer: $P_{n,r}^1 . h_{J,n,r}^2 < P_{th}$,
- $P_{n,r}^2$ the smallest transmit power on ($n$, $r$) that triggers the jammer: $P_{n,r}^2 . h_{J,n,r}^2 > P_{th}$.

These statistics will be used by the BBU for performing RA at the beginning of each new timeslot, as will be shown in Section III.

In NOMA, a user $k_i$ scheduled on subband $n$ can remove the inter-user interference from any other user $k_j$, collocated on $n$, whose channel gain verifies $h_{k_j,n,r} < h_{k_i,n,r}$ [9,10] and treats the received signals from other users as noise. In this study, we consider a maximum number of collocated users per subband of 2, i.e., $m(n) = 1$ or 2. Indeed, it has been shown [10] that the gain in performance obtained with $m(n) = 3$, compared to 2, is minor. Also, limiting the number of multiplexed users per subband reduces the SIC complexity at receivers. We also assume that the same RRH is used to power the signals of the two paired users on a subband. Therefore, in the absence of jamming, two users

$k_1$ and $k_2$ can be paired on subband *n* if their channel gains verify [9, 10]: $h_{k_2,n,r} < h_{k_1,n,r}$. The allocated powers of the two users on *n* must verify the following power multiplexing condition (PMC), in order to allow SIC implementation at $k_1$:
$$P_{k_1,n,r} < P_{k_2,n,r}. \qquad (1)$$

In case of OMA transmission, the binary triggering function of the jammer is given by [7]:
$$\alpha_{n,r} = \begin{cases} 1, & \text{if } P_{k,n,r} h_{J,n,r}^2 > P_{th} \\ 0, & \text{otherwise} \end{cases}$$

In NOMA transmission, it becomes:
$$\alpha_{n,r} = \begin{cases} 1, & \text{if } \left(P_{k_1,n,r} + P_{k_2,n,r}\right) h_{J,n,r}^2 > P_{th} \\ 0, & \text{otherwise} \end{cases}$$

We will denote by first (resp. second) user the one having the higher (resp. lower) channel gain on a subband between the two paired users. Their theoretical throughputs $R_{k_i,n,r}$, $1 \leq i \leq 2$, on *n* depend on the jammer triggering function and can be expressed by the Shannon capacity limit:
$$R_{k_1,n,r} = \frac{B}{S} \log_2 \left( 1 + \frac{P_{k_1,n,r} h_{k_1,n,r}^2}{\sigma^2 + \alpha_{n,r} P_J g_{k_1,n}^2} \right), \qquad (2)$$

$$R_{k_2,n,r} = \frac{B}{S} \log_2 \left( 1 + \frac{P_{k_2,n,r} h_{k_2,n,r}^2}{P_{k_1,n,r} h_{k_2,n,r}^2 + \sigma^2 + \alpha_{n,r} P_J g_{k_2,n}^2} \right), \qquad (3)$$

**Proposition 1.** When the jammer is triggered, user $k_1$ can perform SIC to remove the signal of user $k_2$ if:
$$h_{k_1,n,r} g_{k_2,n} > h_{k_2,n,r} g_{k_1,n} \qquad (4)$$

**Proof:** Let $SINR_{k_2}^{(k_1)}$ the necessary Signal to Interference and Noise Ratio at user $k_1$ to decode the signal of $k_2$:
$$SINR_{k_2}^{(k_1)} = \frac{P_{k_2,n,r} h_{k_1,n,r}^2}{P_{k_1,n,r} h_{k_1,n,r}^2 + \sigma^2 + P_J g_{k_1,n}^2}$$

$k_1$ can perform SIC on *n* if: $SINR_{k_2}^{(k_1)} \geq SINR_{k_2}^{(k_2)}$, where
$$SINR_{k_2}^{(k_2)} = \frac{P_{k_2,n,r} h_{k_2,n,r}^2}{P_{k_1,n,r} h_{k_2,n,r}^2 + \sigma^2 + P_J g_{k_2,n}^2}.$$

By writing $SINR_{k_2}^{(k_1)} - SINR_{k_2}^{(k_2)}$ and using some mathematical calculations, it can be verified that:
$$SINR_{k_2}^{(k_1)} \geq SINR_{k_2}^{(k_2)} \Leftrightarrow \sigma^2 P_{k_2,n,r} \left( h_{k_1,n,r}^2 - h_{k_2,n,r}^2 \right)$$
$$+ P_{k_2,n,r} P_J \left( h_{k_1,n,r}^2 g_{k_2,n}^2 - h_{k_2,n,r}^2 g_{k_1,n}^2 \right) \geq 0.$$

In practical transmission situations, the first term in the above inequality is much smaller than the second, since in general $\sigma^2 \ll P_J g_{k,n}^2$. This directly leads to (4).
By interchanging $k_1$ and $k_2$ in the previous calculations, it can be shown that when (4) is verified, user $k_2$ cannot perform SIC. □

Proposition 1 will be of primary importance for the development of the RA techniques in the NOMA-DAS context. As for the PMC in case of jammer triggering, that allows SIC implementation at $k_1$, it becomes:

$$P_{k_1,n,r} h_{k_1,n,r}^2 + P_J g_{k_1,n}^2 < P_{k_2,n,r} h_{k_1,n,r}^2 \qquad (5)$$

The system EE is expressed as [13]:
$$EE = \frac{\sum_k \sum_n \sum_r R_{k,n,r}}{\varepsilon \left( \sum_k \sum_n \sum_r R_{k,n,r} \right) + \sum_k \sum_n \sum_r P_{k,n,r} + \beta P_{static}}, \qquad (6)$$

where $\varepsilon$ is a coefficient that links the processing power to the overall transmission rate on an antenna (the transmit power per unit throughput), taken as 0.1 W/bps as in [13]. $P_{static}$ is the power needed to operate the circuitry and electronics of one RRH (it is assumed to be independent from the antenna load). $\beta$ is the number of active RRHs (i.e. RRHs having at least one linked user). The second term in the denominator of (6) accounts for the total transmit power on all active RRHs.

In this study, each user is allocated at most one subband in each timeslot. The aim of RA is to jointly determine the optimal power, RRH and subband allocation to each user, so as to maximize the average EE of the system while allowing each user to achieve its target rate over the horizon *Ho*, and by respecting the PMC in case of NOMA transmission. Such problem is a mixed combinatorial and non-convex one. Besides, compared to the case of NOMA CAS transmission, an additional dimension is added to the problem, which is the determination of the best RRH to power each allocated subband to a user. In the sequel, we propose three algorithmic solutions to the problem, two in the OMA context and one in the NOMA context.

### III. RESOURCE ALLOCATION TECHNIQUES FOR THE MAXIMIZATION OF THE ENERGY-EFFICIENCY

#### A. OMA context with single antenna transmission

Our aim is to ensure that all users in the cell achieve on average a rate of $R_{k,target}$ over *Ho*. Since some of these users will not achieve this rate in a certain timeslot due to the jammer presence, in our RA technique, the target rate is re-estimated for each user at the beginning of each timeslot, based on the previous average achieved rate. Let: *i* the timeslot index, $T_S$ the timeslot duration, $R_k^{avg,i}$ the average rate of user *k* until timeslot *i* (i.e. from timeslot 1 until *i* in the Horizon *Ho*), $R_k^{ach,i}$ the achieved rate of *k* in timeslot *i* (only known by the BBU at the end of timeslot *i*), $N_{k,i}$ the number of bits transferred for user *k* over the last *i* timeslots. We can write: $R_k^{avg,i} = N_{k,i} / i T_S$. Also:
$$N_{k,i} = R_k^{avg,i-1}(i-1)T_S + R_k^{ach,i} T_S. \text{ Therefore, we obtain:}$$
$$R_k^{avg,i} = \left(1 - \frac{1}{i}\right) R_k^{avg,i-1} + \left(\frac{1}{i}\right) R_k^{ach,i}. \qquad (7)$$

By setting $R_k^{avg,i} = R_{k,target}$ in (7), the required rate of user *k* at timeslot *i*, $R_k^{req,i}$, is found by:
$$R_k^{req,i} = i R_{k,target} - (i-1). R_k^{avg,i-1}. \qquad (8)$$

It can be verified that if the average rate of *k* over *i* timeslots is less than $R_{k,target}$, the value of the required rate $R_k^{req,i+1}$ will be greater than $R_{k,target}$.

Let $P_k^{avg,i}$ the transmit power for user *k* averaged over timeslots 1 until $i$. Algorithm 1 summarizes the RA technique for OMA, where only a single RRH is used to power a user's signal.

---

*Algorithm 1: MaxEE-OMA*

---

**Initialization**
$i = 1$
$P^1_{n,r}(i) = 0; P^2_{n,r}(i) = \infty$

**Step 1** // User selection
$U_a = \varnothing$ // Set of users with an already allocated subband in the current timeslot
**If** $i = 1$ **then** {
// In the first timeslot, user priority is based on channel gains
**For** each user $k = 1, \ldots, K$
$$h_{k,\max} = \arg\max_{n,r} h_{k,n,r}; \quad h_{k,\text{second}} = \arg\max_{(n,r)/h_{k,n,r} \neq h_{k,\max}} h_{k,n,r}$$
**End for**
$$k^* = \arg\max_{k} \left( h_{k,\max} - h_{k,\text{second}} \right) \}$$
**Else** {
// In subsequent timeslots, user priority is based on power or rate
**If** $K^i = \{k / R_k^{avg,i-1} < R_{k,\text{target}}\} = \varnothing$ **then** $k^* = \arg\max_{k} P_k^{avg,i-1}$
**Else** $k^* = \arg\max_{k \in K^i} \left( R_{k,\text{target}} - R_k^{avg,i-1} \right) \}$

**Step 2** // Subband selection for user $k^*$
**For** each available subband
// Find the needed power $\forall (n,r)$ supposing first that J is not triggered
$$P_{k^*,n,r} = \frac{\sigma^2}{h^2_{k^*,n,r}} \left( 2^{S \cdot R^{req,i}_{k^*}/B} - 1 \right)$$
**If** $P_{k^*,n,r} < P^1_{n,r}(i)$ **then** $R^{avg,i}_{k^*} = R^{req,i}_{k^*}$ // J will not be triggered at the end of timeslot $i$
**If** $P_{k^*,n,r} > P^2_{n,r}(i)$ **then** $P_{k^*,n,r} = \frac{\sigma^2 + P_J g^2_{k^*,n}}{h^2_{k^*,n,r}} \left( 2^{S \cdot R^{req,i}_{k^*}/B} - 1 \right)$ // J will be triggered and power is recalculated accordingly
**If**
$$\left( P^1_{n,r}(i) < P_{k^*,n,r} < P^2_{n,r}(i) \right) \& \left( \left| P^1_{n,r}(i) - P_{k^*,n,r} \right| < \left| P^2_{n,r}(i) - P_{k^*,n,r} \right| \right)$$
**then**
take $P_{k^*,n,r} = P^1_{n,r}(i)$ // Conservatory mode → $R^{avg,i}_{k^*} < R^{req,i}_{k^*}$
**If**
$$\left( P^1_{n,r}(i) < P_{k^*,n,r} < P^2_{n,r}(i) \right) \& \left( \left| P^2_{n,r}(i) - P_{k^*,n,r} \right| < \left| P^1_{n,r}(i) - P_{k^*,n,r} \right| \right)$$
**then** $P_{k^*,n,r} = \frac{\sigma^2 + P_J g^2_{k^*,n}}{h^2_{k^*,n,r}} \left( 2^{S \cdot R^{req,i}_{k^*}/B} - 1 \right)$ // Aggressive mode →

$R^{avg,i}_{k^*} = R^{req,i}_{k^*}$ if J is triggered, $R^{avg,i}_{k^*} > R^{req,i}_{k^*}$ if J not triggered
**End for**

**If** $G^i_{k^*} = \{(n,r) / R_{k^*,n,r} = R^{req,i}_{k^*}\} = \varnothing$ **then**
$(n^*, r^*) = \arg\max_{(n,r)} R_{k^*,n,r}$ // If no subband allows $k^*$ to achieve $R^{req,i}_{k^*}$, retain the subband yielding the highest rate.

**Else**
$(n^*, r^*) = \arg\max_{(n,r)} EE,$

$$EE = \frac{R_{U_a} + R_{k^*,n,r}}{\varepsilon \left( R_{U_a} + R_{k^*,n,r} \right) + \left( \sum_{k' \in U_a} \sum_{n'} \sum_{r'} P_{k',n',r'} + P_{k^*,n,r} \right) + \beta P_{static}},$$

where $R_{U_a} = \sum_{k' \in U_a} \sum_{n'} \sum_{r'} R_{k',n',r'}$.
$U_a = U_a \cup \{k^*\}$.

Repeat Steps 1 and 2 until Card$\{U_a\} = K$

**Step 3** // Update jammer statistics $\forall (n, r)$ at the end of timeslot $i$
**If** $\alpha_{n,r}(i) = 1$ and $P_{k,n,r} < P^2_{n,r}(i)$ **then** $P^2_{n,r}(i+1) = P_{k,n,r}$
**If** $\alpha_{n,r}(i) = 0$ and $P_{k,n,r} > P^1_{n,r}(i)$ **then** $P^1_{n,r}(i+1) = P_{k,n,r}$

$i = i + 1$
Repeat Steps 1 to 3 until $i = Ho$

---

### B. NOMA transmission

In Algorithm 2, NOMA is applied as a subsequent stage to single-user resource allocation. Let $S_{\text{sole}}$ the set of subbands with a single allocated user and $k_2$ the user selected for NOMA allocation. $S_{\text{NOMA}}(k_2)$ is the set of candidate subbands considered for the assignment of $k_2$ as second user. The first user already assigned to a certain subband $n$ is designated by $k_1(n)$. The RRH powering the signal of user $k_1$ on $n$ is used for the transmission of the signal of $k_2$ on $n$.

The PMCs are tested at each stage of the NOMA allocation algorithm, and the second user power is increased when necessary to fulfill the PMC with a gap $m$ that depends on SIC practical implementation. In this work, a value of $m = 0.01$ W was taken.

Note that when a user's power is increased to respect the PMC, the achievable rate of this user on its allocated subband is increased accordingly. Therefore, such user may exceed its requested rate at the end of the current timeslot. This excess is compensated for at the subsequent timeslot using (8).

---

*Algorithm 2: MaxEE-NOMA*

---

**At each timeslot**
Perform Steps 1 and 2 from Algorithm 1
$S_{sole} = \{1, 2, \ldots, S\}$

**Step 3**
$k_2 = \arg\max_{k \in K^i} \left( R_{k,\text{target}} - R_k^{avg,i} \right)$ // Identify the user with the furthest distance from its target rate
$S_{\text{NOMA}}(k_2) = \varnothing$
$\partial R_{k_2} = R^{req,i}_{k_2} - R_{k_2,n,r}$ // Rate deficit of user $k_2$
// $R_{k_2,n,r}$ is the expected rate of $k_2$ at the end of the OMA phase

**For** each subband $n$ in $S_{sole}$

// Constitute the set of candidate NOMA subbands for the pairing of $k_2$

**If** $\alpha_{n,r}(i) = 1$ **then** { // $J$ is considered to be triggered on $n$ based on the prior assignment of $k_1(n)$

**If** $h_{k_1,n,r} \cdot g_{k_2,n} > h_{k_2,n,r} \cdot g_{k_1,n}$ **then** {

$S_{\text{NOMA}}(k_2) = S_{\text{NOMA}}(k_2) \cup \{n\}$; $S_{\text{sole}} = S_{\text{sole}} \cap \{n\}^C$
// Find the power needed to fill the rate deficit of $k_2$

$p_{k_2,n,r} = \left(2^{\partial R_{k_2} \cdot S/B} - 1\right) \cdot \left(p_{k_1,n,r} \cdot h_{k_2,n,r}^2 + \sigma^2 + P_J g_{k_2,n}^2\right) / h_{k_2,n,r}^2$ }

**If** $p_{k_1,n,r} \cdot h_{k_1,n,r}^2 + P_J g_{k_1,n}^2 > p_{k_2,n,r} \cdot h_{k_1,n,r}^2$ **then** // Verify the PMC

$p_{k_2,n,r} = (1+m) \cdot \dfrac{p_{k_1,n,r} \cdot h_{k_1,n,r}^2 + P_J g_{k_1,n}^2}{h_{k_1,n,r}^2}$ }

**If** $\alpha_{n,r}(i) = 0$ **then** // $J$ is considered not triggered based on $k_1(n)$
{
**If** $h_{k_1,n,r} > h_{k_2,n,r}$ **then** {

$S_{\text{NOMA}}(k_2) = S_{\text{NOMA}}(k_2) \cup \{n\}$; $S_{\text{sole}} = S_{\text{sole}} \cap \{n\}^C$

$p_{k_2,n,r} = \left(2^{\partial R_{k_2} \cdot S/B} - 1\right) \cdot \left(p_{k_1,n,r} \cdot h_{k_2,n,r}^2 + \sigma^2\right) / h_{k_2,n,r}^2$ }

**If** $p_{k_1,n,r} > p_{k_2,n,r}$ **then** $p_{k_2,n,r} = (1+m) p_{k_1,n,r}$

// Verify if, while the assignment of $k_1$ did not trigger the jammer, the latter is triggered due to the power increase on subband $n$ incurred by NOMA multiplexing

**If** $p_{k_1,n,r} + p_{k_2,n,r} < P_{n,r}^1(i)$ **then** $p_{k_1,n,r}$ and $p_{k_2,n,r}$ are unchanged

**If** $p_{k_1,n,r} + p_{k_2,n,r} > P_{n,r}^2(i)$ **then** { // Jammer will be triggered
**If** $h_{k_1,n,r} \cdot g_{k_2,n} < h_{k_2,n,r} \cdot g_{k_1,n}$ **then**
// $n$ is not an appropriate candidate for pairing $k_2$
$S_{\text{NOMA}}(k_2) = S_{\text{NOMA}}(k_2) \cap \{n\}^C$; $S_{\text{sole}} = S_{\text{sole}} \cup \{n\}$
**Else** {

$p_{k_1,n,r} = \left(2^{R_{k_1,n,r}^{OMA} \cdot S/B} - 1\right) \cdot \left(\sigma^2 + P_J g_{k_1,n}^2\right) / h_{k_1,n,r}^2$ // re-estimate the power of $k_1$ so as to maintain its rate from the OMA phase $R_{k_1,n,r}^{OMA}$

$p_{k_2,n,r} = \left(2^{\partial R_{k_2} \cdot S/B} - 1\right) \cdot \left(p_{k_1,n,r} \cdot h_{k_2,n,r}^2 + \sigma^2 + P_J g_{k_2,n}^2\right) / h_{k_2,n,r}^2$ }

**If** $p_{k_1,n,r} \cdot h_{k_1,n,r}^2 + P_J g_{k_1,n}^2 > p_{k_2,n,r} \cdot h_{k_1,n,r}^2$ **then**

$p_{k_2,n,r} = (1+m) \cdot \dfrac{p_{k_1,n,r} \cdot h_{k_1,n,r}^2 + P_J g_{k_1,n}^2}{h_{k_1,n,r}^2}$ }

**If** $P_{n,r}^1(i) < p_{k_1,n,r} + p_{k_2,n,r} < P_{n,r}^2(i)$ and $p_{k_1,n,r} + p_{k_2,n,r}$ closer to $P_{n,r}^1(i)$ than to $P_{n,r}^2(i)$ **then** Reduce $p_{k_2,n,r}$ so that $p_{k_1,n,r} + p_{k_2,n,r} = P_{n,r}^1(i)$

**If** $P_{n,r}^1(i) < p_{k_1,n,r} + p_{k_2,n,r} < P_{n,r}^2(i)$ and $p_{k_1,n,r} + p_{k_2,n,r}$ closer to $P_{n,r}^2(i)$ **then** apply the same calculations as in the case where $p_{k_1,n,r} + p_{k_2,n,r} > P_{n,r}^2(i)$
}
**End for**

**If** $S_{\text{NOMA}}(k_2) = \varnothing$ **then** $k_2$ is no longer considered for NOMA pairing at subsequent pairing stages

**Else** {
**If** $J_{k_2}^i = \{n / R_{k_2,n,r} \geq \partial R_{k_2}\} = \varnothing$ **then** $n^* = \arg\max_n R_{k_2,n,r}$
**Else** $n^* = \arg\max_n EE$ }

Repeat Step 3 of Algorithm 2 until $S_{\text{sole}} = \varnothing$ or no more users can be paired

Update jammer statistics as in Step 3 of Algorithm 1

---

### C. OMA with multiple antenna transmission

When a user $k$ is in outage (i.e. did not reach its target rate), instead of using NOMA to pair $k$ on one or more subbands as done in Algorithm 2, another possibility is to fill its rate deficit by selecting a second antenna to transmit on the same subband assigned to $k$. This approach will be referred to as MAT (multi-antenna transmission). In MAT, relying on the statistics of $P_{n,r}^1(i)$ and $P_{n,r}^2(i)$ is no longer practical because the jammer, instead of detecting the power $p_{k,n,r} \cdot h_{J,n,r}^2$, will now detect the composite signal $p_{k,n,r_1} \cdot h_{J,n,r_1}^2 + p_{k,n,r_2} \cdot h_{J,n,r_2}^2$ where $r_1$ and $r_2$ are the two RRHs transmitting the signal of user $k$. Since two different RRHs are involved in the same signal transmission, the system is unable to predict the triggering outcome at the end of the timeslot and will instead consider the worst case scenario (i.e. that the jammer is triggered).

Let $k$ be a user in outage in timeslot $i$, $n$ its assigned subband, $r_1$ the antenna to which the user was originally connected and $r_2$ an additional candidate antenna for connecting $k$. The achievable rate of $k$ is now:

$$R_{k,n,(r_1,r_2)} = \dfrac{B}{S} \log_2\left(1 + \dfrac{p_{k,n,r_1} \cdot h_{k,n,r_1}^2 + p_{k,n,r_2} \cdot h_{k,n,r_2}^2}{P_J \cdot g_{k,n}^2 + \sigma^2}\right).$$

By letting $R_{k,n,(r_1,r_2)} = R_k^{req,i}$, we can determine the power needed on $r_2$ as:

$$p_{k,n,r_2} = \dfrac{\left(2^{R_k^{req,i} \cdot S/B} - 1\right)\left(\sigma^2 + P_J \cdot g_{k,n}^2\right) - p_{k,n,r_1} \cdot h_{k,n,r_1}^2}{h_{k,n,r}^2} \quad (9)$$

---

### Algorithm 3: MaxEE-OMA-MAT

Perform Steps 1 and 2 from Algorithm 1

**For** each user $k$ in outage
    **For** each candidate RRH $r_2 \neq r_1(k)$
    Find $p_{k,n,r_2}$ using (9)
    Calculate the system EE
    **End for**
$r_2^* = \arg\max_{r_2}(EE)$
**End for**

Update jammer statistics

## IV. SIMULATION RESULTS

The proposed RA techniques are simulated in the LTE/LTE-Advanced context [14]. We adopt a hexagonal cell model with a radius $R_d$ of 500 m. In the CAS case ($R = 1$), a unique RRH is positioned at the cell center. In DAS, a number $R$ of RRHs of 4 or 7 was considered. In each case, one antenna is located at the cell center, while the others are equally distanced and positioned on a circle of radius $2R_d/3$ centered at the cell center. The transmission medium is a frequency-selective Rayleigh fading channel with a root mean square delay spread of 500 ns, distance-dependent path loss with decay factor 3.76 and lognormal shadowing with 8 dB variance. We also take $B = 10$ MHz, $S = K = 16$, $N_0 = 4.10^{-18}$ mW/Hz, and $Ho = 10$ timeslots, with $T_S = 1$ ms. The jammer triggering threshold $P_{th}$ is $10^{-12}$ W; it is unknown to the BBU and used only to assess the performance of the algorithms based on the jamming triggering outcome.

We start first, in Fig. 2, by showing the average total transmit power in the cell, in terms of $R_{k,target}$, for the different methods. Focusing on OMA transmission, it is clear how the increase in the number of RRHs decreases the necessary power. The gain of DAS with $R = 7$, towards CAS, can reach 50 dBW at $R_{k,target} = 5$ Mbps. Now considering $R = 7$, MaxEE-OMA-MAT presents the worst performance compared to the two other methods, since it aims at filling the rate gap of users in outage, at the expense of the transmit power increase. This is also the case of MaxEE-NOMA which, for the same reasons, is outperformed by MaxEE-OMA, until the target rate surpasses 8 Mbps. At higher values of $R_{k,target}$, the power performance is in favor of MaxEE-NOMA, because of the significant reduction of the requested rates at later timeslots within $Ho$.

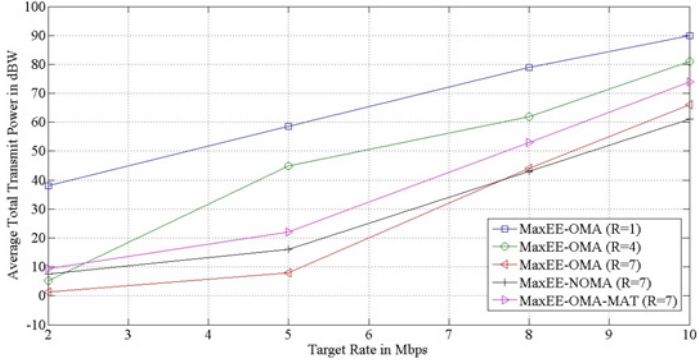

Fig. 2. Average total transmit power in terms of $R_{k,target}$, for $P_{static} = 1$ W and $P_J = 1$ W.

However, even at low values of $R_{k,target}$, Fig. 3 shows that the maximum EE is achieved by NOMA transmission, which allows a better usage of the available spectrum (i.e. achieving higher data rates), even when the transmit power is higher. Also, as the jamming power $P_J$ increases, the EE decreases for all three methods as a result of the decrease in the achieved rate of the user transmissions that trigger the jammer.

Fig. 4 shows the influence of $P_{static}$ on the system behavior. It is clear how the increase of the antenna static power decreases the EE for all methods, which can be explained by Eqt. (6).

In Fig. 5, one can see that the number of active RRHs (i.e. RRHs that have at least one connected user) decreases when the static power needed to operate an antenna increases. Even though the total transmit power increases when the number of involved RRHs is reduced, the BBU tries to select an RRH that has already connected users, when assigning a new subband to a selected user, in order to avoid wasting circuit power over one user on a new RRH. However, if the transmit power cost is higher than the static power cost, the user is connected to the suitable antenna that maximizes the EE. This can be further observed in Fig. 6 that shows the increase of the transmit power with $P_{static}$, due to the decrease in the number of active RRHs. Note that the latter is the same for MaxEE-NOMA and MaxEE-OMA, since MaxEE-NOMA uses power multiplexing on already assigned subbands to increase user rates, without the need for involving additional RRHs, which is not the case for MaxEE-OMA-MAT.

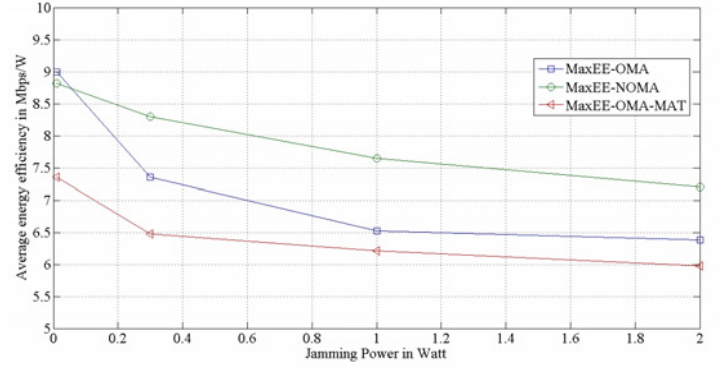

Fig. 3. Average energy efficiency in terms of the jamming power for $R = 7$, $P_{static} = 1$ W, and $R_{k,target} = 5$ Mbps.

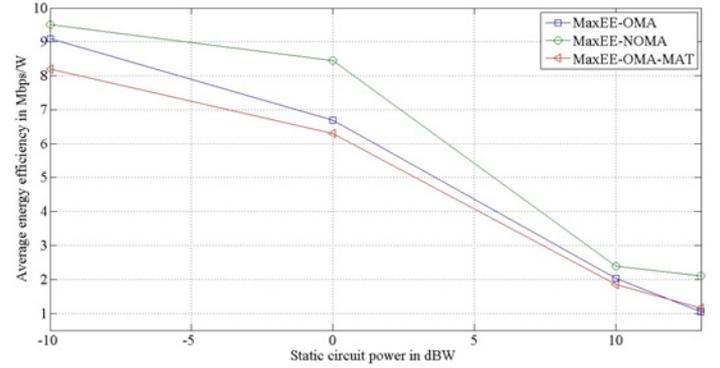

Fig. 4. Average energy efficiency in terms of $P_{static}$ for $R = 7$, $P_J = 1$ W and $R_{k,target} = 5$ Mbps.

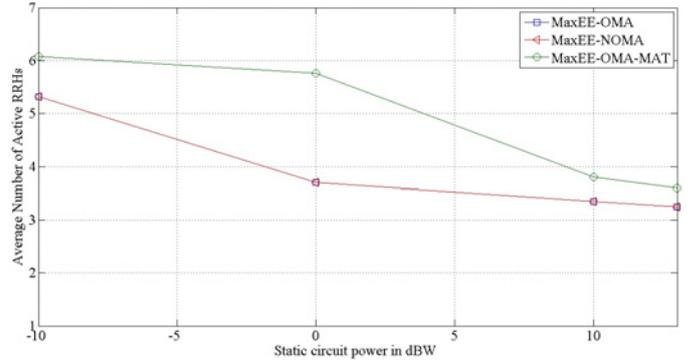

Fig. 5. Average number of active RRHs in terms of $P_{static}$ for $R = 7$, $P_J = 1$ W and $R_{k,target} = 5$ Mbps.

Fig. 7 shows that the percentage of users that perform NOMA slightly increases as $P_J$ increases. On the one hand, the increase in $P_J$ reduces the rate for user transmissions that trigger the jammer, thus necessitating more NOMA pairings to fulfill rate gaps. On the other hand, as $P_J$ increases, stricter constraints (e.g.

(4) and (5)) are imposed on eligible users for allowing pairings. In general, even for $P_{static}$ = 10 dB, no more than 12.3 % of the users perform NOMA, which limits the average SIC complexity at the receiver level.

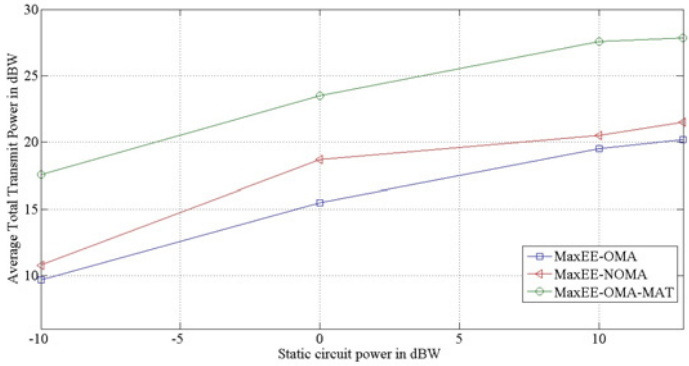

Fig. 6. Average total transmit power in terms of $P_{static}$ for $R$ = 7, $P_J$ = 1 W and $R_{k,target}$ = 5 Mbps.

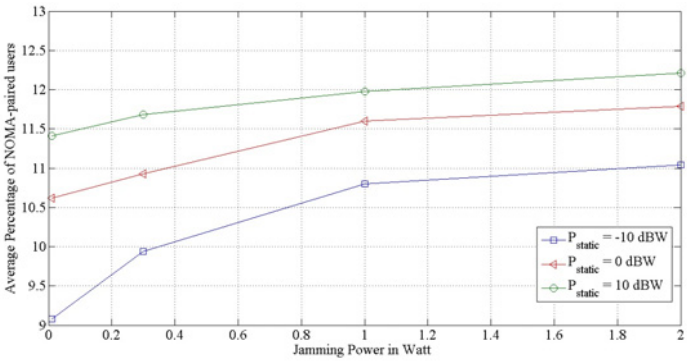

Fig. 7. Average percentage (%) of NOMA-paired users in MaxEE-NOMA in terms of the jamming power for $R$ = 7 and $R_{k,target}$ = 5 Mbps.

In Fig. 8, we can see that the average percentage of users in outage generally decreases over the horizon. MaxEE-OMA-MAT presents the lowest percentage (which quickly reaches 0), since it focuses on filling the rate gaps using additional antenna transmissions, at the cost of an increase in the transmit power, inherent to the worst-case scenario considered in Algorithm 3. However, with MaxEE-NOMA (resp. MaxEE-OMA), no more than 5 % (resp. 10 %) of the users fail to reach their goal at the end of the time horizon.

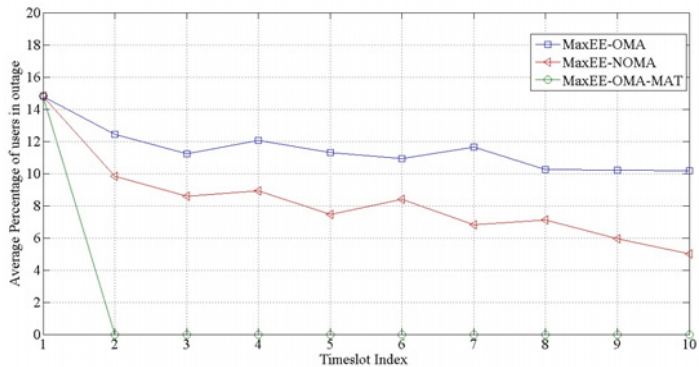

Fig. 8. Average percentage (%) of users in outage over the Horizon $H_o$ for $R$ = 7, $P_J$ = 1 W, $P_{static}$ = 1 W and $R_{k,target}$ = 5 Mbps.

## V. CONCLUSION

In this work, we introduced three new techniques for optimizing the system EE in the presence of a reactive jammer. Based on the past observations of the jammer behavior, and using the concepts of DAS and NOMA, the system exploits bandwidth and power resources so as to allow the maximum number of users to achieve a target rate. The combination of DAS with NOMA shows the best EE performance over OMA and/or CAS, as well as with respect to multi-antenna transmission, which increases the transmit power and the number of active RRHs. The study could be further extended to allow multi-channel allocation to each user, through elaborate PA techniques that take the jammer characteristics into account.

## VI. ACKNOWLEDGEMENT

This work has been funded with support from the Lebanese University. We also thank the ELSAT2020 project.